\documentclass[twocolumn,amsmath,amssymb,pra,showpacs]{revtex4}

\usepackage[T1]{fontenc}
\usepackage[latin1]{inputenc}
\usepackage{graphicx}
\usepackage{natbib}

\begin{document}
\title{Scattering and absorption of ultracold atoms by nanotubes}
\author{$^1$B. Jetter, $^1$J. Märkle, $^{1,2}$P. Schneeweiss, $^1$M. Gierling, $^3$S. Scheel, $^1$A. G\"unther, $^1$J. Fort\'agh, $^1$T.~E. Judd}
\affiliation{$^1$CQ Center for Collective Quantum Phenomena and their Applications in LISA+, Physikalisches Institut, Eberhard-Karls-Universität Tübingen, Auf der Morgenstelle 14, D-72076 Tübingen, Germany}
\affiliation{$^2$TU Wien, Atominstitut, Stadionallee 2, 1020 Wien, Austria}
\affiliation{$^3$Institut für Physik, Universität Rostock, Universitätsplatz 3, D-18051 Rostock, Germany}

\date{\today}

\begin{abstract}
We investigate theoretically how cold atoms, including Bose-Einstein condensates, are scattered from, or absorbed by nanotubes with a view to analysing recent experiments. In particular we consider the role of potential strength, quantum reflection, atomic interactions and tube vibrations on atom loss rates. Lifshitz theory calculations deliver a significantly stronger scattering potential than that found in experiment and we discuss possible reasons for this. We find that the scattering potential for dielectric tubes can be calculated to a good approximation using a modified pairwise summation approach, which is efficient and easily extendable to arbitrary geometries. Quantum reflection of atoms from a nanotube may become a significant factor at low temperatures, especially for non-metallic tubes. Interatomic interactions are shown to increase the rate at which atoms are lost to the nanotube and lead to non-trivial dynamics. Thermal nanotube vibrations do not significantly increase loss rates or reduce condensate fractions, but lower frequency oscillations can dramatically heat the cloud.
\end{abstract}


\maketitle

\section{Introduction}

Interfacing ultracold atom clouds with micro- and nanoscaled objects opens the door for creating hybrid quantum systems which combine the long quantum coherence times of ultracold atoms with the flexibility of solid state technologies \cite{RevModPhys.79.235,PhysRevLett.107.223001,Stehle2011}. There have been studies of cold atoms with ions and superconductors \cite{Zipkes2010,PhysRevLett.105.133202,1367-2630-12-6-065024}, proposals for hybrid nanosystems \cite{chang2010,PhysRevLett.102.033003,salem2010} and recent experiments in which the interaction between cold atoms and carbon nanotubes was observed and measured \cite{GierlingM.2011,SchneeweissP.2012}.

However, there is much to learn about how atoms scatter from nanostructures, or are absorbed by them. Experiments typically measure loss rates from atom clouds to solid structures but these can be influenced by a number of factors including dispersion potential strengths, elastic reflection, electromagnetic spin flips, interatomic interactions and possible mechanical vibrations of the nanostructures. The issue of spin-flips has been quantitatively assessed \cite{PhysRevA.75.062905} but the other factors require further work.
Progress on these issues has been stalled by the fact that calculating dispersion potentials (which generally dominate the interaction) in non-trivial geometries requires very heavy numerics. In spite of work on the problem since the 1930s \cite{PhysRevA.80.012504} full calculations using Lifshitz theory have only recently become possible. Therefore, in addition to comparing the full theory with experiment and establishing the relevant physics, there is an additional need to explore approximate alternatives for calculating dispersion potentials.

Scattering of cold atoms on small objects may be analysed using non-dynamical approaches \cite{Zipkes2010} and quantum scattering theories \cite{PhysRevA.81.062714,0295-5075-102-3-33001}, but recent experiments with cold atoms and nanotubes have left open questions. In particular, there is a lack of theory calculations with which to compare the measured potential strength.
Better understanding of the dispersion interactions and scattering in such systems is crucial to future hybrid experiments and will aid the design of nanotechnologies where such forces exert a profound influence on their mechanical properties. It may also provide insight into the manufacture of nanostructures since this is often done from the vapour phase. 

In this paper, we study how cold atoms are either elastically scattered from, or absorbed by, a nanotube. We consider both thermal clouds and Bose-Einstein condensates. In particular, we explore whether and when physical effects such as quantum reflection, interatomic interactions and thermal tube vibrations are important to loss rates from cold atom clouds. We begin by calculating the Casimir-Polder (CP) potentials \cite{footnotenamingcon} for conducting and insulating tubes in Section 2. We find that the calculated potentials do not match those seen in experiment and discuss reasons for this. Quantum reflection may play a role here, so in Section 3 we use the potentials to study the likelihood of elastic scattering by calculating quantum reflection probabilities for a range of incident velocities. We then look at the role of interatomic interactions by performing dynamical simulations of a Bose-Einstein condensate (BEC) interacting with a nanotube in Section 4, to determine the rate at which atoms are lost due to inelastic scattering. We find that interatomic interactions increase the loss rates, but that the potential strength has little effect on loss rates in this case. Finally, in Section 5, we consider how an oscillating nanotube can affect the loss rates and the condensate fraction of a cold atom cloud when the two are overlapped, again by performing time-dependent simulations.

\section{Casimir-Polder Potential}

The interaction between an atom and a cylindrical, electrically neutral object is expected to be dominated by dispersion forces, specifically the Casimir-Polder potential. This subject has been considered by several groups \cite{barash,PhysRevB.75.235413,PhysRevB.65.155402,PhysRevA.80.012504,Fink2011,PhysRevB.76.045417} and is complicated by the fact that the properties of the cylinder alter not only the strength coefficients, but also the functional form of the potential \cite{barash}. The potential in the case of an atom and an infinitely long tube can be calculated using the general Lifshitz formula \cite{parsegianvdw,PhysRevB.72.035451,PhysRevA.75.062905}
\begin{equation}
\label{eq:scheeltubevdw}
V_{CP}(r)=\mu_0 k_B T \sum_{l=0}^{\infty}{}' \xi_l^2 \alpha(i\xi_l) \mathrm{Tr}\left[\mathbf{G}^{(s)}(r,r,i\xi_l)\right]
\end{equation}
where $r=\sqrt{x^2+y^2}$ is the separation between the atom (in the $x$-$y$ plane) and the tube axis $z$, $T$ is the temperature of the tube, $\xi_l=2\pi k_B T l/\hbar$ are the Matsubara frequencies with mode number $l$ (the dash on the sum implies half weight for the first term), $\alpha(i\xi_l)$ is the polarizability of the atom, and the trace is taken over $\mathbf{G}^{(s)}(r,r,i\xi_l)$, the scattering part of the Green's polarizability tensor for the tube. The other symbols have their usual meaning.

Equation (\ref{eq:scheeltubevdw}) is appropriate for conducting and insulating tubes, but the potential in the insulating case  can be approximately modelled by using the simpler pairwise summation approach \cite{parsegianvdw}
\begin{equation}
\label{eq:pwsvdw}
V_{CP}(r)=-\int_V \frac{C(r,\xi_l)}{r_{pws}^6}\:dx\:dy\:dz
\end{equation}
where $r_{pws}$ is the separation between the atom and an infinitesimal piece of the tube, $C(r,\xi_l)$ is the strength coefficient, and the integral takes place over the volume of the tube, $V$. It is possible to calculate $C$ as a function of $r$ and $\xi_l$ to provide approximate corrections to the potential due to retardation and temperature effects. 
We use the general expression for the dispersion interaction between two point particles \cite{parsegianvdw}
\begin{eqnarray}
\label{eq:Cpars}
C(r,\xi_l) = &\: 6k_B T \sum_{l=0}^{\infty}{}'  \frac{\alpha(i\xi_l)\beta(i\xi_l)}{(4\pi\epsilon_0)^2} \times \nonumber \\
&\left(1+r_l+\frac{5}{12}r_l^2+\frac{1}{12}r_l^3+\frac{1}{48}r_l^4\right) e^{-r_l}
\end{eqnarray}
where $r_l=(2\xi_l/c)r_{pws}$ with $c$ the vacuum speed of light and $\beta(i\xi_l)$ is the polarizability of a discrete piece of nanotube (derived from the bulk dielectric constant) used for the pairwise summation. As $\alpha(i\xi_l)$ generally decays much more rapidly with frequency than $\beta(i\xi_l)$, we can use static tube polarizabilities to a good approximation.
\begin{figure}
\includegraphics[width=0.9\columnwidth]{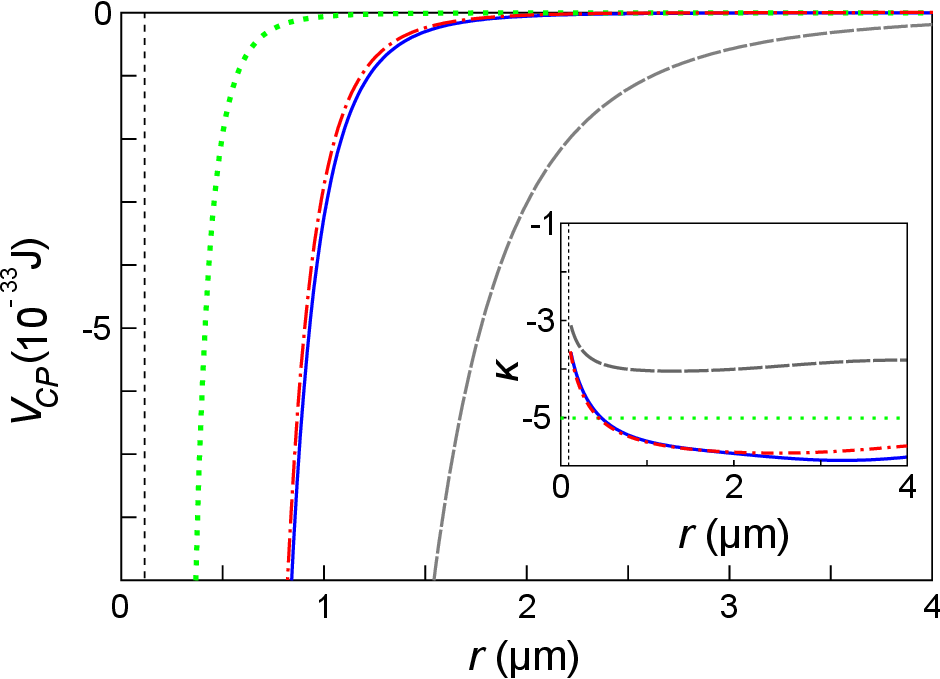}
\caption{\label{fig:vntube}(a) CP potential energy as a function of $r$ for a $^{87}$Rb atom and a nanotube. Conducting tube, Lifshitz model (grey dashed curve), dielectric tube, Lifshitz model (blue solid curve) and the dielectric tube, pairwise model (red dot-dash curve). The green dotted curve shows the fitted model $-C_5/r^5$ from experiment. Inset: effective power law exponent, $\kappa$, for the four cases, colour scheme as in the main plot. The vertical dashed lines mark the position of the nanotube surface.}
\end{figure}

We plot the potential for $^{87}$Rb atoms interacting with a solid nanotube at $T=300\:$K in Fig.\ \ref{fig:vntube} assuming a tube radius  $R_0=63.5\:$nm, based on recent experiments \cite{SchneeweissP.2012}. The $^{87}$Rb polarizability is taken from \cite{PhysRevA.69.022509}. We plot one curve (grey, dashed) using Eq.\ (\ref{eq:scheeltubevdw}) and the Green's tensor from \cite{PhysRevA.75.062905}, which is for a good conductor. 
The blue curve is also calculated using Eq.\ (\ref{eq:scheeltubevdw}) but assumes a dielectric tube with relative permittivity $\epsilon = 2.5$, a good insulator.  The Green's tensor coefficients are obtained from \cite{Li:2000:0920-5071:961} in this case.
The dotted red curve is plotted using Eqs.\ (\ref{eq:pwsvdw}) and (\ref{eq:Cpars}) with a tube length $L=10\:\mu$m and $\beta$ calculated using Clausius-Mossotti relations, assuming a dielectric constant $\epsilon=2.5$ and a number density of $10^{29}$, typical for carbon systems. The potential is calculated at $z=5\:\mu$m, directly between the two ends of the tube.
 
The conducting tube generally delivers the strongest potential, as expected; the dielectric tube potential is around two orders of magnitude weaker. The curves for the dielectric tube using the Lifshitz and pairwise sum approaches are very similar, the discrepancy being $\sim 10\%$ at $r=0.5\:\mu$m. This close match is significant since Eq.\ (\ref{eq:pwsvdw}), using Eq.\ (\ref{eq:Cpars}), can typically be calculated orders of magnitude faster than Eq.\ (\ref{eq:scheeltubevdw}) and is immediately extendable to arbitrary geometries. A Born expansion of the Casimir-Polder potential confirms that these theories should be close \cite{buhmannborn}. 

It is instructive to consider how the effective power law exponents of the full theory vary, depending on the atom-tube separation and the model used. The inset of Fig.\ \ref{fig:vntube} shows these exponents for the curves, $\kappa(r) = d({\rm{log}}\:\left|V_{CP}\right (r)|)/d({\rm{log}}\: r)$, plotted against $r$. The grey dashed curve (conducting tube) goes to $\kappa=-3$ at the surface and to $\kappa \approx -4$ at $r \sim 2\:\mu$m. By contrast, the dielectric model (blue curve) is closer to $\kappa = -5$. The pairwise model potential is for a finite tube and ultimately goes to $\kappa = -6$ for $r \gg L = 10\:\mu$m and $r \gg hc/k_B T \approx 8\:\mu$m, the thermal wavelength, as expected.

For comparison we show the model $-C_5/r^5$ potential \cite{footnoter5} fitted to the experimental results with $C_5=6\times 10^{-65}\:$J$\:$m$^5$ [green dashed curve]. This value was obtained by fitting the thermal cloud results to a classical scattering theory \cite{SchneeweissP.2012}. We see that even the dielectric potential is much stronger than that obtained in experiment \cite{SchneeweissP.2012}. For example, at $r=300\:$nm the fitted potential returns $V=2.5\times10^{-32}\:$J whereas the Lifshitz theory for the dielectric predicts $V=1.2\times10^{-29}\:$J. In spite of the large error bars on the $C_5$ coefficient, which suggest $C_5$ could be up to an order of magnitude larger, theory and experiment are far from agreement. We now discuss possible reasons for this.

A small extracted value for the potential energy implies low loss rates of atoms from the cloud to the tube. However, technical problems with hybrid experiments (spin flips, technical heating, patch potentials) tend to increase loss rates. Exceptions to this might occur because of calibration problems when positioning the atoms (so that only low density regions of the cloud overlap with the nanotube) or quantum reflection of atoms from the tube. If the classical fit routine is performed assuming a uniform reflection probability of 50\% for all atoms, the extracted $C_5$ coefficient rises by a factor of $\sim 60$. However, in the next section, we show that few atoms are likely to reflect with this probability. A more detailed analysis requires a large range of calculations for different velocities and potentials and is beyond the scope of this paper.

Uncertainties in the horizontal positioning can only account for up to a 15\% change in the loss rates. Likewise, uncertainties in the vertical separation between surface and cloud are expected to be submicron and even shifts of $\sim 2\mu$m do not lead to sufficiently large changes in the extracted $C_5$ coefficient.

A common side-effect of carbon nanotube growth is the deposition of large quantities of amorphous carbon \cite{huang:460} and electron micrographs of the tubes in the experiment suggest this was also the case here, resulting in a more insulating tube.
Furthermore, this phenomenon leads to rough surfaces \cite{huang:460}. On the one hand surface roughness is expected to increase the potential between an atom and a surface \cite{PhysRevA.61.022115}. Using a pairwise approximation, corrections of up to 70\% may be expected.  However, quantum reflection experiments by Pasquini et al.\ \cite{PhysRevLett.97.093201} used a pillared surface which reduced the density of the surface by two orders of magnitudes. Making the ansatz that the Casimir-Polder potential was weakened by the same factor, they obtained agreement between theory and experiment for their reflection results.  

We expect effects due to finite tube length and possible variations in the polarizability per atom to be small by comparison. Without further details of the material properties involved, this discussion must remain speculative but we believe that a reduced surface density due to roughness, combined with reflection effects, may provide the largest contribution to the discrepancy. It is worth pointing out that attempts to quantum reflect cold atoms from aerogels (expected to have a similar potential strength to that obtained in the experiments, as well as similar roughness) observed no reflection in spite of theoretical expectations that the probabilities would lie above 70\% \cite{PhysRevLett.97.093201}. Reasons for this also remain speculative. This all suggests that further work is needed to understand Casimir-Polder potentials due to rough structures, particularly nanostructures. 

As a final point in this discussion, we note that obtaining precise values for the potential strength through atom losses is difficult because of the potential steepness. Capture radii can be obtained with reasonable accuracy but errors are amplified by the large power laws when attempting to use this data to calculate potential strengths.

\section{Quantum Reflection}

Elastic quantum reflection is one of the processes that can reduce loss rates to the nanotube and is also interesting in its own right, giving information about the tube's scattering properties. We now analyze quantum reflection of atoms from the tube using the potentials obtained in the previous section. Our aim is to assess what difference the potential strengths make, whether reflection probabilities are significant and measureable, and to what extent results for BECs and thermal clouds are affected by it.

Quantum reflection is a phenomenon that allows particles to reflect from a scattering potential in the absence of a classical turning point, e.g.\ a sharp potential drop. The condition for strong quantum reflection is given by $\Phi(k)=(1/k^2)dk/dr \sim d\lambda/dr \gtrsim 1$ \cite{shimizu} where the local wave number, $k=2\pi/\lambda$, depends on the de Broglie wavelength, $\lambda$, for the particle's centre-of-mass motion. The potentials we consider here are purely attractive so this is the only elastic scattering channel \cite{1367-2630-13-8-083020}. 

For simplicity, we restrict our analysis to the component of incident velocity that is perpendicular to the surface, $v_{perp}$, and reduce the problem to one dimension (1D). The quantum reflection probabilities can then be calculated with 1D Schrödinger wavepackets in a harmonic trap \cite{Cornish2009}. This also allows us to unambiguously distinguish quantum reflection effects from small angle scattering. Atoms that do not quantum reflect are lost, either by adsorption on the tube or inelastic scattering. We model these losses with an imaginary potential \cite{footnoteqrcond,1367-2630-12-6-063033}. 

\begin{figure}
\includegraphics[width=0.9\columnwidth]{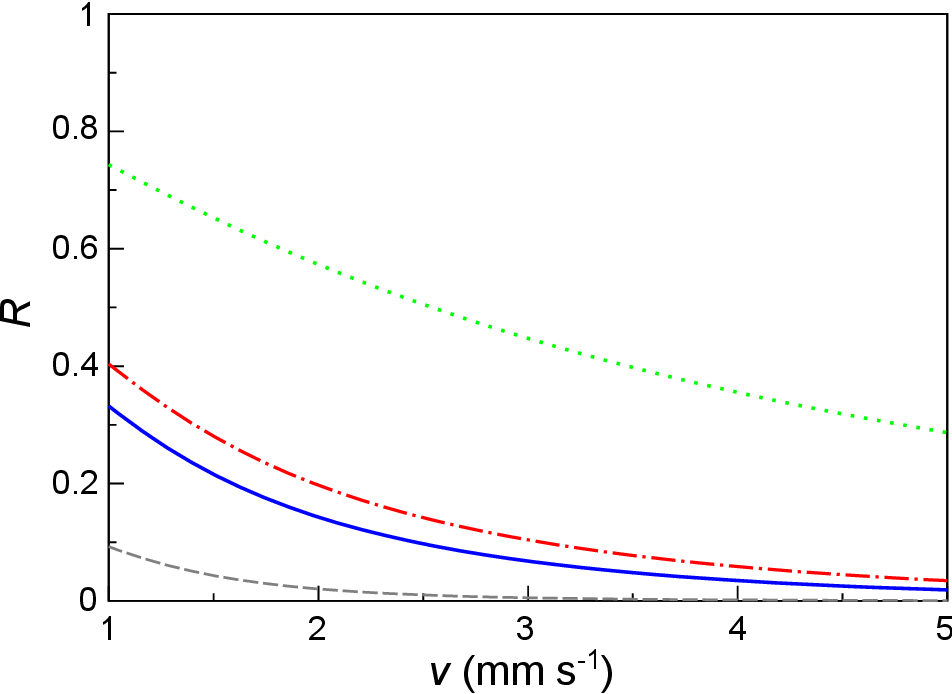}
\caption{\label{fig:qr}Quantum reflection probabilities, $R$, plotted against perpendicular incident velocity, for the different nanotube potentials. Conducting tube, Lifshitz model (grey dashed curve), dielectric tube, Lifshitz model (blue solid curve), dielectric tube, pairwise model (dot-dash red curve), and the fitted potential from experiment (green dotted curve, see text).}
\end{figure}

The reflection probabilities, $R$, for the four different potentials are plotted in Fig.\ \ref{fig:qr} for perpendicular incident velocities in the range $v_{perp}=1 \ldots 5\:$mm$\:$s$^{-1}$ which are common to ultracold atom clouds.
We see that the conducting tube (grey dashed curve) barely reflects $^{87}$Rb atoms at the relevant velocities, which is expected for stronger potentials \cite{PhysRevLett.97.093201}. The dielectric tube reflects atoms better (Lifshitz and pairwise models, solid blue and dot-dash red curves respectively), delivering probabilities above $\sim 0.2$ when the velocities drop below $\sim 2\:$mm$\:$s$^{-1}$. This corresponds to the mean velocity expected for a thermal cloud with $T\sim 40\:$nK. The weak $-C_5/r^5$ potential from experiment (green dotted curve) delivers noticeably higher reflection probabilities that are still above 25\% at $v \sim 5\:$mm$\:$s$^{-1}$.

The results suggest that quantum reflection may affect loss rates for thermal clouds at low temperatures and should certainly influence atom scattering if a BEC is made to interact with a nanotube. The reflection probabilities for dielectric tubes are large enough to produce measurable changes in the atom loss rates for thermal clouds, but not enough to explain the large discrepancy between theory and experiment for the potential.

\section{Interatomic Interactions}

Quantum hybrid systems are potentially at their most powerful when using quantum coherent atoms such as those in a BEC. However, the behaviour of BECs is often strongly affected by interatomic interactions. In addition to thermal clouds, the experiments of Schneeweiss et al. \cite{SchneeweissP.2012} considered scattering of a $^{87}$Rb BEC from a nanotube; we now perform a theoretical analysis of this system to ascertain how BEC-nanotube interactions and loss rates differ from the thermal case. 

We use the conducting Lifshitz and $-C_5/r^5$ potentials, and assume cylindrical symmetry with the BEC fully overlapped with the tube (the dielectric case has been omitted for clarity). We model the dynamics of the condensate using the time-dependent Gross-Pitaevskii equation (GPE) in three dimensions \cite{pethick}
\begin{equation}
\label{eq:gpe}
-\frac{\hbar^2}{2m}\nabla^2\Psi+V_{E}\Psi+\frac{4\pi\hbar^{2}a}{m}\left|\Psi\right|^2\Psi=i\hbar\frac{\partial\Psi}{\partial t}
\end{equation}
where $a=5.4\:$nm is the $s$-wave scattering length \cite{pethick}, $m$ is the atomic mass, $\nabla^2$ is the Laplacian in cylindrical coordinates, and $\Psi(z,r,t)$ is the wave function at time $t$, normalized such that $\left|\Psi\right|^2$ is the number of atoms per unit volume. The external potential, $V_E$, is a combination of the Casimir-Polder potential and the harmonic trap used to confine the atom cloud.  As with the reflection simulations, we employ imaginary potentials to absorb atoms that do not reflect. We choose trap frequencies $\omega_z=2\pi \times 80\:$rads$\:$s$^{-1}$ and $\omega_r=2\pi \times 37\:$rads$\:$s$^{-1}$ in the vertical and horizontal (radial) directions respectively, which are close to the experiments \cite{GierlingM.2011,SchneeweissP.2012}. The initial atom number is chosen by averaging the first two points of the experimental curve we wish to compare with. Spin flip losses are expected to be negligible \cite{PhysRevA.75.062905}.

\begin{figure}
\includegraphics[width=0.9\columnwidth]{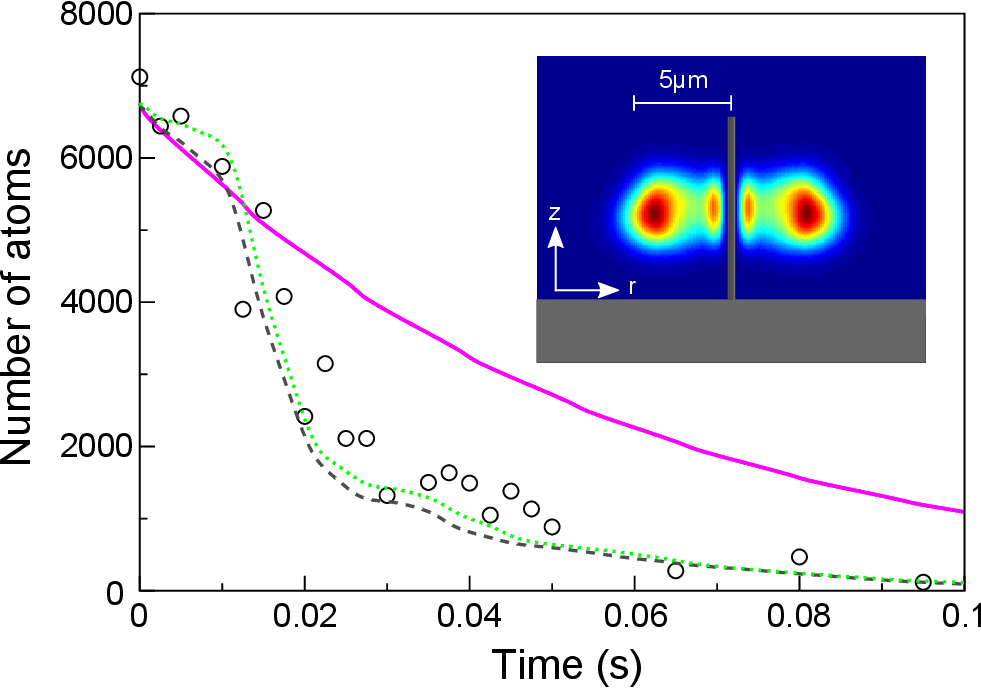}
\caption{\label{fig:becloss}BEC losses due to spatial overlap with a nanotube, number of atoms plotted against time. Black circles are experimental data. The grey dashed curve and the green dotted curve are for GPE simulations with the Lifshitz conducting and $C_5$ potentials respectively (see text). The magenta curve is for a non-interacting (Schrödinger) wavepacket. Inset shows the cross-section of the BEC's density profile after 0.03 seconds showing non-trivial dynamics. Grey shapes sketch the relative position of the tube and the chip surface. White arrows show the axes and the white bar shows scale.
}
\end{figure}

We plot the number of atoms in the BEC against time in Fig.\ \ref{fig:becloss} for the GPE simulation with the Lifshitz conducting potential (grey dashed curve), the $-C_5/r^5$ potential (green dotted curve) and the corresponding experimental points (black circles).
The simulated loss curves are non-trivial showing a wide range of loss rates. The first portion of the curve, where the signal-to-noise ratio is higher, may be be approximated by a simple exponential curve, as was done in experiment \cite{SchneeweissP.2012}. However, this more sophisticated analysis reveals additional effects.

We see that the strength of the potential has only a small effect on the loss rates for BECs, in contrast to what is expected for thermal clouds \cite{SchneeweissP.2012}. This is expected because in all cases considered here, the de Broglie wavelength of the atoms ($\sim 10\:\mu$m) is much larger than the size of the region in which they experience a strong potential ($\sim 1\:\mu$m); the potential can be approximately modelled as a sharp step \cite{PhysRevLett.95.073201} (this is only possible in the case of BECs, not thermal clouds). As has been shown in previous studies, atoms are lost due to the curvature of the wavefunction, which creates a quantum pressure that forces atoms towards the absorbing surface \cite{Cornish2009,PhysRevA.77.043603}. Hence BECs appear to be poor probes of the details of nanostructures, the physics being dominated by the BEC's properties, rather than the nanostructures'.

Analysis of the simulation dynamics reveals that atoms are accelerated towards the tube and the slow initial loss rate increases as atoms are rapidly removed from the BEC. This generates collective oscillations in the trap, shown in the inset of Fig.\ \ref{fig:becloss}, along with sound waves and phase gradients that can produce interference patterns. The subsequent fast loss rate flattens into a plateau; at this point very few atoms are near the trap centre due to the collective oscillations. Finally for $t\gtrsim 0.05\:$s we see a slower loss rate; at this point most of the atoms have been absorbed so the repulsive interatomic interactions, and hence loss rates, are lower. 

For comparison, we also show a simulation with the dielectric potential and no interatomic interactions [Fig.\ \ref{fig:becloss}, magenta curve]. All other parameters remain the same. We see a simple exponential decay, with a loss rate that is significantly slower than in the experiment, revealing the importance of atomic interactions \cite{Cornish2009}.

\section{Nanotube Oscillations}

Thus far, we have assumed the nanotube to be a static object. However, it is possible that the interaction between atoms and nanotubes is significantly affected by thermally excited mechanical oscillations of the nanotube. We therefore end our investigation by considering the effects of a vibrating nanotube on a BEC with a simple model system. In addition to ascertaining the role of thermal efffects in the experiments (not to mention general nanosystems), this also affords us the opportunity to consider the effect of a vibrating cantilever on cold atom systems, which is also the subject of current work \cite{PhysRevLett.107.223001}.

Here we model a full three-dimensional system (no cylindrical symmetry as in the previous section) with the potential further simplified to a step function with side lengths $x=470\:$nm and $z=390\:$nm. The tube is aligned along the $y$-axis in this case \cite{footnotecapturerad}. In the light of the previous section and Ref.\ \cite{PhysRevLett.95.073201}, the results using this simplified potential and a BEC are not expected to differ greatly from those using a full CP potential. The dynamics are again modelled by the GPE with $10^4$ $^{87}$Rb atoms, trap frequencies as before. The simulation begins with the BEC ground state in the trap. The tube potential is then ramped up over $1\:$ms which simulates insertion. The tube is then oscillated sinusoidally for $0.02\:$s in the $x$-direction with an amplitude of $A$, frequency $f$. Spatial and energy cut-offs absorb atoms that are excited to very high energies by the process; we typically lose 80-90\% of the atoms. The tube potential is then suddenly turned off and the one-particle density matrix $\rho^{(1)}(z,z')=\left\langle \Psi^\dagger(z)\Psi(z') \right\rangle$ of the cloud is evaluated. This is done using the ergodic hypothesis \cite{PhysRevA.72.063608}, which allow us to replace the ensemble average by a time average of the line $\Psi(0,0,z)$ over 0.04 seconds. This enables us to estimate the relative occupation of the ground state mode, $f_c$, after Penrose-Onsager \cite{PhysRev.104.576}.

\begin{figure}
\includegraphics[width=0.9\columnwidth]{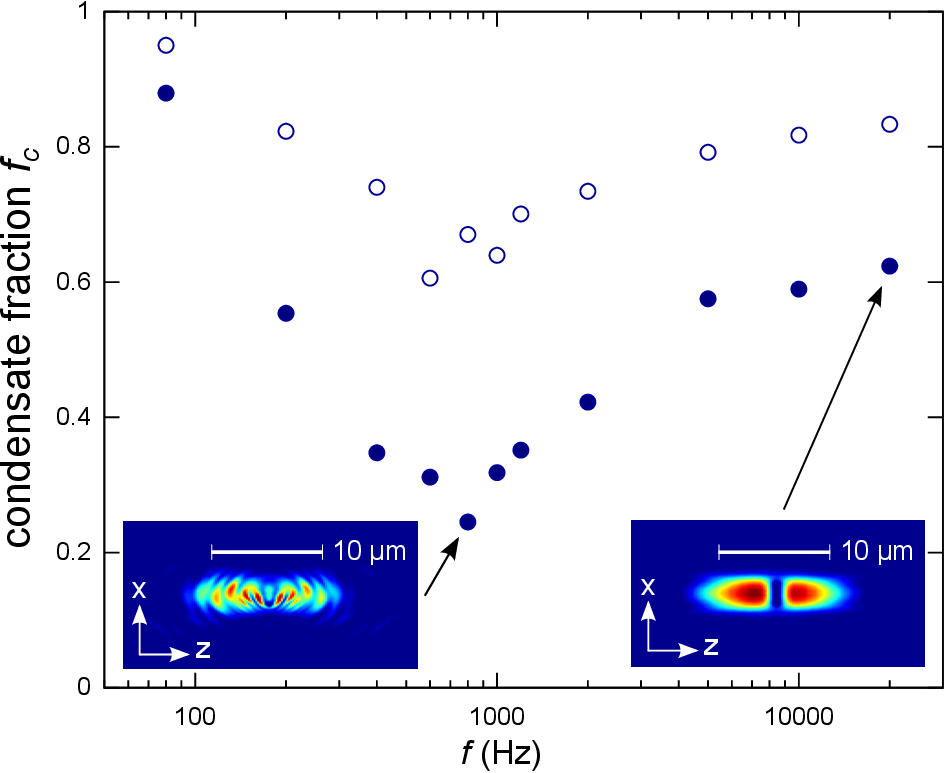}
\caption{\label{fig:osctube}Simulated points show the remaining condensate fraction against nanotube oscillation frequency following a $0.02\:$s overlap between a BEC and a nanotube with oscillation amplitude $A=0.75\:\mu$m (open points) and $A=1.5\:\mu$m (solid points). Insets show cross-section cloud profiles at the end of the oscillation process. Arrows indicate axes, white bars give scale.}
\end{figure}

We plot $f_c$ against the vibration frequency of the tube in Fig.\ \ref{fig:osctube}.
At low frequencies $f_c$ is largely maintained. However, at higher frequencies, the tube oscillations significantly reduce $f_c$. The simulated points contain fluctuations because particular frequencies excite trap resonances more strongly than others. At even higher frequencies, $f_c$ becomes larger again, saturating at a constant value ($f_c$ does not return exactly to 1.0 since the insertion and removal of the tube leaves excitations in the cloud). Unsurprisingly, larger amplitudes create more disruption.

The high energy of the high frequency oscillations might be expected to strongly disrupt the nonlinear system, but we do not observe this. The behaviour is explained by the BEC's speed of sound, $v_s$, and correlation time, $t_c=l_H/v_s$ where $l_H$ is the condensate's healing length \cite{pethick}. The lower frequency limit for the formation of topological excitations and the reduction of $f_c$ occurs where the speed of the tube first exceeds the local speed of sound, $A \cdot 2\pi f \gtrsim v_s$. This corresponds to $f \gtrsim 60\:$Hz for $A=1.5\:\mu$m and the population of excited modes increases with the frequency. However, if the tube oscillates with a frequency $f \gg 1/t_c$, the gas cannot respond to the oscillations and merely sees a static, time-averaged potential; the disruption is minimal, and $f_c$ is retained [Fig. \ref{fig:osctube}, inset bottom right]. In our case, $1/t_c \sim 3000\:$Hz, assuming the peak atom density. Between these limits, the BEC is strongly disrupted [Fig. \ref{fig:osctube}, inset bottom left].

Thermally excited mechanical vibrations in nanotubes typically have frequencies $f > 100\:$kHz and amplitudes $\ll 1\:\mu$m \cite{PhysRevB.58.14013}. We therefore conclude that such vibrations are unlikely to increase atom losses or disrupt the BEC. The operation of hybrid devices consisting of ultracold atoms and nanostructures should not be impaired by these effects.

\section{Conclusion}

In conclusion, we studied a range of factors that might influence loss rates from cold atom clouds to a nanotube. We found that potential strengths obtained in experiment were much weaker than those predicted by Lifshitz theory and further work will be required to settle this issue.  Surface roughness and quantum reflection may be important factors. Pairwise theories of dispersion interactions are not generally additive \cite{PhysRevA.82.052517} and only guaranteed to become accurate in the limit of rarified media, but we find a close match with the Lifshitz theory in the dielectric case; this could potentially make design of nanodevices simpler. 

We studied quantum reflection from nanotubes and showed it becomes significant for $^{87}$Rb at low temperatures. This is in contrast to a rubidium atom and a wall, where reflection is expected to be barely measurable. Quantum reflection appears to play a greater role than previously expected but cannot fully explain the very low loss rates observed in experiment by itself. 

We then simulated BEC atom losses to a nanotube, finding highly non-trivial dynamics; repulsive interatomic interactions were found to increase atom loss rates. In contrast to what is expected for a thermal cloud, BEC loss rates were largely insensitive to the potential form and strength. 

Finally we considered how an oscillating nanotube interacts with a BEC, concluding that thermal vibrations will not change loss rates from an atom cloud. For driven oscillations, we found there is a resonance where the condensate fraction is significantly reduced at a frequency $< 1/t_c$. Very high and very low frequencies damage the condensate much less. In the future it may be possible to explore how the cold atoms exert a back-action on nanostructures, possibly leading to cooling since cold atom clouds typically have similar masses to nanotubes. This will probably have to been done with levitated nanodevices, or millikelvin temperature surfaces.

We gratefully acknowledge support from the DFG through SFB/TRR21 and the Open Access Publishing Fund of Tübingen University, the Baden-Württemberg RiSC Programme and ``Kompetenznetz Funktionelle Nanostrukturen'', the Carl Zeiss Stiftung, and BW-grid computing resources.

\providecommand{\newblock}{}


\begin{thebibliography}{10}
\bibitem{footnotenamingcon}
 We use the naming convention that Casimir-Polder potentials are between a
  polarizable particle and a solid object as opposed to van der Waals
  potentials (between particles/molecules) and Casimir potentials (between
  macroscopic objects) \cite{scheelrev}.
\bibitem{footnoter5}
 This potential form is the non-retarded limit for an infinitely long and thin
  insulating tube \cite{barash}.
\bibitem{footnoteqrcond}
 The quantum reflection condition, $\Phi(k)$, defines a quantum reflection
  zone, which typically lies between $r\sim60\:$nm and $r\sim2\:\mu$m for this
  system. Atoms which come closer than the inner boundary cannot reflect and
  are expected to be lost to the tube. At this point we therefore cut off the
  potential. Beyond this point, an imaginary potential is introduced, starting
  from a value of zero at the cut-off and increasing towards the surface
  \cite{PhysRevA.78.053623}. The exact strength and form of the function used
  is irrelevant provided all atoms are absorbed without triggering anomalous
  reflection.
\bibitem{footnotecapturerad}
 The atom capture radius for a nanotube is expected to be a few hundred
  nanometres. The chosen widths are numerically convenient; the exact values
  should not be crucial to the results. The change of tube axis, so that it
  oscillates in the plane with tighter trapping, is also a numerical
  convenience, making it easier to achieve threshold resolution.
\end{thebibliography}
\end{document}